# Nonrelativistic spin-splitting multiferroic antiferromagnet and compensated ferrimagnet with zero net magnetization

Jianting Dong[1†], Kun Wu[1†], Meng Zhu[1], Fanxing Zheng[1], Xinlu Li[1], and Jia Zhang[1*]

[1]*School of Physics and Wuhan National High Magnetic Field Center, Huazhong University of Science and Technology, 430074 Wuhan, China*

[*] jiazhang@hust.edu.cn
[†]Authors contributed equally to this work.

## Abstract

Spin-splitting antiferromagnets with spin-polarized band structures in momentum space have garnered intensive research attention due to their zero net magnetic moments, ultras fast spin dynamics as conventional antiferromagnets, and spin-polarized transport properties akin to ferromagnets, making them promising candidates for antiferromagnetic spintronics. However, unlike spin-torque switching of ferromagnets by electric current, efficient electric control of spin-splitting antiferromagnetic order remains challenges. In this work, we identify prototypes of multiferroic spin-splitting antiferromagnets, including $BiFeO_3$, $Fe_2Mo_3O_8$ and compensated ferrimagnet $GaFeO_3$ with ferroelectric polarization as well as spin-polarized electronic structures. We establish design principles for the spin-splitting multiferroic antiferromagnets and compensated ferrimagnets, elucidating the band symmetry features in Brillouin zone. We demonstrate that the spin polarization in spin-splitting magnets, despite of zero net magnetic moment, can be switched by ferroelectric polarization, providing an efficient means of controlling the antiferromagnetic order. Our work may inspire future development of novel multiferroic functional magnets with zero magnetic moments and pave the way for their applications in magnetoelectric spintronic devices.

# Introduction

Recently discovered nonrelativistic spin-splitting antiferromagnets (AFMs), belonging to a new type of magnetism, characterized by zero net magnetic moment like conventional spin-degenerate AFMs, yet exhibiting spin-polarized band structures in momentum space even in the absence of spin-orbit coupling (SOC), akin to ferromagnets[1][2][3][4]. This unique property stems from the breaking of $PT\tau$ symmetry (a combination of spatial inversion $P$, time reversal $T$ and fractional translation $\tau$) and $U\tau$ symmetry ($U$ is the spinor symmetry related to spin reversal), which can manifest in both collinear and noncollinear antiferromagnetic spin orders. Tetragonal $RuO_2$, $MnF_2$ and hexagonal CrSb *etc.*, also named as altermagnets, are representative collinear spin-splitting AFMs, which possess certain rotational operations connecting opposite spin sublattices[5][6][7]. Typical noncollinear spin-splitting AFMs include the $D_{019}$ phase $Mn_3X$ (X=Sn, Ge, Ga)[8], cubic phase $Mn_3Pt$[9], $Mn_3Ir$ *etc*. Both collinear and noncollinear spin-splitting AFMs are shown to have intriguing transport phenomena, such as the anomalous Hall, unconventional spin Hall and magnetoresistance effect *etc*, making them promising for advancing antiferromagnetic spintronics[9][10][11][12][13][14]. Additionally, while most spin-splitting AFMs show spin polarization along particular paths in Brillouin Zone, a distinct class of compensated magnet (actually compensated ferrimagnets) has been claimed with spin-polarized band structures throughout the entire Brillouin Zone[5].

In parallel, ferroelectrics require the breaking of $P$ symmetry with empty *d/f* orbitals in transition metals, while ferromagnets arise from the violation of time reversal symmetry and originate in partially filled *d/f* orbits in transition metals[15]. The combination of ferroelectricity and magnetism creates the field of multiferroics[16], enabling energy-efficient control of magnetic properties via electric fields. Research on multiferroics has flourished since the discoveries of antiferromagnetic $BiFeO_3$[17] and $TbMnO_3$[18]. Subsequently, multiferroic ferromagnets and ferrimagnets have also been identified.

Examples of multiferroic ferromagnets includes strained $EuTiO_3$[19][20] and the 2D material

VOF$_2$[21]. Multiferroic ferrimagnets are represented by Fe$_3$O$_4$[22], frustrated antiferroelectric BaFe$_{12}$O$_{19}$[23], spin spiral CoCr$_2$O$_4$[24], double-perovskite Er$_2$CoMnO$_6$[25] and noncollinear ferrimagnets such as YbFeO$_3$[26] and CaBaCo$_4$O$_7$[27][28][29]. Meanwhile, besides BiFeO$_3$ and TbMnO$_3$, multiferroic AFM also include helical magnet NiI$_2$[30] and MnWO$_3$[31], collinear magnetoelectric Cr$_2$O$_3$[32] and noncollinear Ba$_2$XGe$_2$O$_7$ (X = Co, Mn, Cu)[33][34][35]. The prosperity of multiferroic materials has significantly broadened the applications of magnetoelectric coupling. However, ferromagnets and uncompensated ferrimagnets, with their net magnetic moments, suffer from stray fields between neighboring cells and are susceptible to disturbances from external magnetic fields, limiting their potential application in high-density spintronic devices. Conventional multiferroic antiferromagnets, on the other hand, exhibit spin degenerate over the entire Brillouin zone, complicating the detection of the Néel vector during magnetoelectric coupling. As a result, efforts to control magnetism via electric fields have primarily limited to multiferroic ferromagnets and ferrimagnets with net magnetic moment[36][37][38].

In this work, we delve into the fundamental principles governing a new class of multiferroics: spin-splitting multiferroic antiferromagnet (SSP-MAFM) and multiferroic compensated ferrimagnet (MCFIM). By employing spin-structure motif pairs and spin space groups (SSGs), we identify various spin-splitting paths in momentum space of SSP-MAFM BiFeO$_3$ and Fe$_2$Mo$_3$O$_8$, as well as MCFIM GaFeO$_3$. Finally, by unveiling these design principles, we pave the way for the development of novel multiferroic AFM with tailored functionalities, offering potential for future spintronic and magnonic applications leveraged by electric fields.

## Computational methods

The first-principles calculations are performed by the Vienna *ab* initio simulation package (VASP)[39][40] with the projector augmented wave (PAW) pseudopotential[41] and the Perdew-Burke-Ernzerhof (PBE) type of the generalized gradient approximation (GGA) of the exchange correlation potential[42]. To have accurate description on the ground state, the Dudarev type of Hubbard U[43] is introduced with U$_{eff}$ = 4 eV for Fe, Mo in Fe$_2$Mo$_3$O$_8$ and Fe in GaFeO$_3$. The Monkhorst–Pack[44] $\boldsymbol{k}$-meshes for self-consistence are 15 × 15 × 6 for $BiFeO_3$, 10 × 10 × 6 for $Fe_2Mo_3O_8$ and 17 × 10 × 10 for $GaFeO_3$, respectively. The atom positions are fully optimized until the force on each ion is less than $10^{-3}$ eV/Å. Since we are discussing the nonrelativistic spin-splitting of antiferromagnet and compensated ferrimagnet, we do not take into spin-orbit coupling (SOC) effect and therefore, the magnetic moments are treated as collinear spin structure. We consider the $BiFeO_3$ ground AFM state of G-type, while that of $Fe_2Mo_3O_8$ and $GaFeO_3$ are C-type and A-type, respectively.

## RESULTS AND DISCUSSIONS

### A. Symmetry analysis

As shown in Fig. 1(a), conventional AFM such as $L1_0$-Mn$X$ ($X$ = Pt, Pd, and Ir)[45] and NiO exhibit nonrelativistic spin-degenerate band structures due to the protection of $PT\tau$ or $U\tau$ symmetry. As magnetic states grow more intricate, in Fig. 1(b), centrosymmetric AFMs break both $PT\tau$ and $U\tau$ symmetries leading to spin-polarized electronic band structures in momentum space, which are typically represented by tetragonal $RuO_2$, hexagonal CrSb and $D0_{19}$ $Mn_3X$ ($X$ = Sn, Ga, Ge), cubic $Mn_3Pt$, $Mn_3Ir$ and antiperovskite $Mn_3GaN$. However, such AFMs cannot exhibit ferroelectricity due to their inherent spatial inversion symmetry.

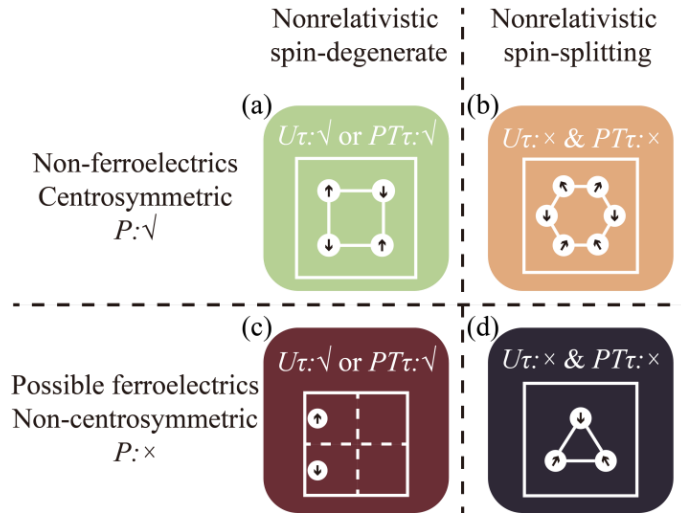

FIG. 1. Classification of AFMs according to three principles ($P$, $PT\tau$ and $U\tau$).

(a) Conventional AFMs with neither ferroelectric polarization nor spin splitting such as $L1_0$-Mn$X$ ($X$ = Pt, Pd, and Ir) and NiO *etc*. (b) Centrosymmetric spin-splitting AFMs like tetragonal $RuO_2$, hexagonal CrSb and $D0_{19}$ Mn$_3X$ ($X$ = Sn, Ga, Ge). (c) Possible multiferroic AFM with spin-degenerate band structures, including multiferroic $Cr_2O_3$, $Ba_2MnGe_2O_7$. (d) Possible multiferroic spin-splitting AFM represented by $BiFeO_3$ and $Fe_2Mo_3O_8$. The two-dimensional representative of different types of AFMs and multiferroic AFMs is adopted for exhibition. The white squares represent the unit cell while the arrows imply magnetic moments.

Multiferroic offers an opportunity to control magnetic states by electric field, where the ferromagnetism and ferroelectricity are coupled by Dzyaloshinskii–Moriya interaction (DMI) in certain non-centrosymmetric systems[15]. Magnetoelectric or ferroelectric AFMs must break spatial inversion symmetry. However, non-centrosymmetric AFMs for instance CuMnAs are not necessarily magnetoelectric or ferroelectric[46]. As depicted in Fig. 1(c), non-centrosymmetric AFM without spatial inversion symmetry $P$ may exhibit magnetoelectric or ferroelectric properties, such as $Cr_2O_3$[32] and $Ba_2MnGe_2O_7$[34][35]. Nevertheless, they are not spin-splitting AFMs due to the preservation of $U\tau$ symmetry in $Ba_2MnGe_2O_7$ and $PT\tau$ symmetry in $Cr_2O_3$, respectively.

As illustrated in Fig. 1(d), antiferromagnets that simultaneously break all three symmetric operations, i.e., $P$, $PT\tau$ and $U\tau$, such as $Ba_2CoGe_2O_7$[34][35], $BiFeO_3$[47] and $Fe_2Mo_3O_8$[48] and so forth, may exhibit multiferroic with spin-splitting band structures. This unique combination implies potential electric field manipulation of antiferromagnetic states, electronic band structures and spin dependent transport properties, *etc*.

## B. Multiferroic spin-splitting antiferromagnet

### I. BiFeO$_3$

As one of the representative multiferroic, $BiFeO_3$ has a high Curie temperature ($T_C$) about 1100 K for its ferroelectric order and a Néel temperature ($T_N$) above 640 K for its antiferromagnetic order[49][50][51][52][53]. The ferroelectrically controlled magnon spin transport has been demonstrated in $BiFeO_3$[54][55]. The lattice constants of $BiFeO_3$ in hexagonal lattice, demonstrated in Fig. 2(a), are $a=b=5.63$ Å and $c=13.87$ Å[47], with magnetic space group of $Cc'$. The principal symmetries of BiFeO$_3$ can be explicitly figured out by spin-structure motif pairs[56][57][58] as illustrated in Fig. 2(b), where two Fe sublattices (indicated by red and blue arrows) are surrounded by octahedral oxygen atoms. It's easy to confirm that $PT\tau$, $U\tau$ and $P$ symmetries are simultaneously broken in BiFeO$_3$, leading to the spin-splitting band structures shown in Fig. 2(g-i). In fact, spin-structure motif pairs fill the gap where crystallographic structural motifs are not sufficient to describe AFM with zero net magnetization[56][57][58][59][60][61]. Meanwhile, as illustrated in Fig. 2(d), the two opposite spin sublattices are related by a 180º rotation operation, classifying BiFeO$_3$ as a special spin-splitting AFM, *i.e.*, altermagnet.

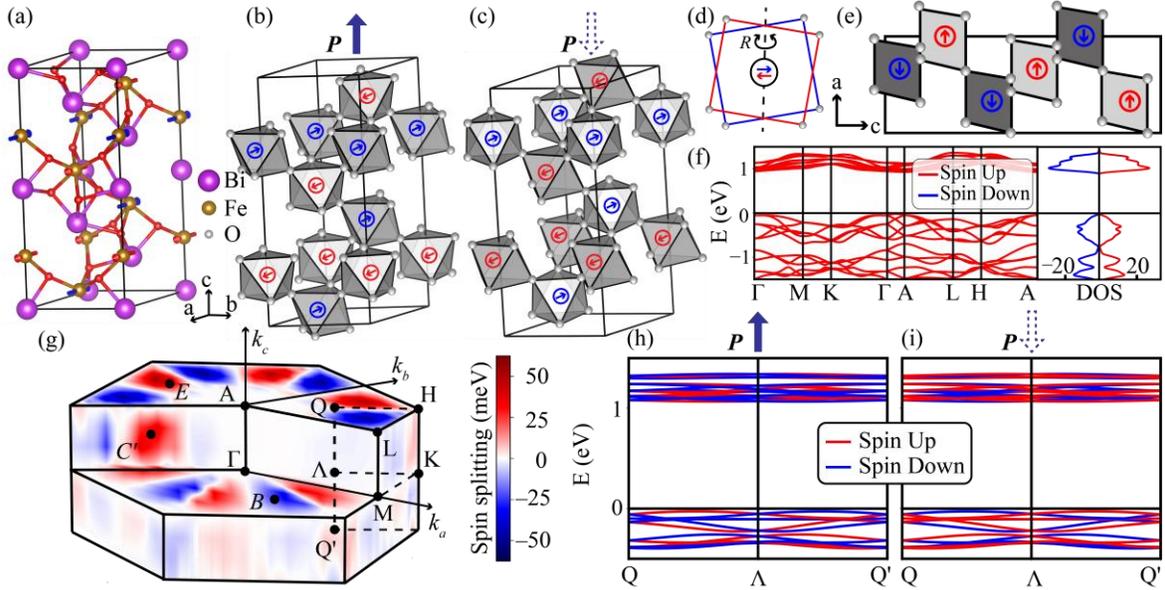

FIG. 2. (a) Atomic and spin structure of hexagonal BiFeO$_3$. The red and blue arrows indicate the opposite magnetic moments on Fe atoms. (b-c) Spin structure motifs of BiFeO$_3$ for ferroelectric polarization up and down states. (d) One of the BiFeO$_3$ motif pairs is connected by a 180º rotation operation. (e) Spin-structure motifs from the $[\bar{1}2\bar{1}0]$ perspective of (b). (f) Density of state (DOS) and band structure along high-symmetry paths. (g) Spin-splitting energy dispersion in the whole Brillouin zone. The splitting energy is calculated from the top valence band. (h-i) Spin-polarized band structure along $k_z$ direction for two ferroelectric states of BiFeO$_3$.

By using first principles calculations, we analyze the spin-splitting band structures of BiFeO$_3$

in its hexagonal lattice. The calculated ferroelectric polarization by Berry phase method is 91.40 $\mu C/cm^2$ and the band gap is 0.63 eV, both agree well with previous results[62][63][64][65][66]. As demonstrated in Fig. 2(e), BiFeO$_3$ in [$\bar{1}2\bar{1}0$] perspective shows $U\tau$ symmetry, resulting in spin degenerate band structures along this high symmetry path shown in Fig. 2(f). Further analysis confirms that the electron band structures are spin degenerate along all high symmetry paths in Brillouin zone, as shown in Fig. 2(g). Especially, at Γ point, the bands do not show spin-splitting due to the presence of a 180° rotation operation connecting two opposite spin motif pairs. Meanwhile, the density of states (DOS) plotted in Fig. 2(f) is identical for spin up and spin down channels, which is a hallmark characteristic of altermagnets.

The Band structures of BiFeO$_3$ are spin-splitting along lower symmetry paths in the Brillouin zone. As illustrated in Fig. 2(g), the color map indicates the spin-splitting energy of the spin up and spin down bands, with the notch exhibiting degenerate paths. Beyond the representation of spin motif pairs, the reversed splitting energies along $k_z$, more clearly illustrated in Fig. 2(h), and the three-fold symmetries can also be obtained from the little groups within the SSGs framework[67]. As demonstrated in Fig. 2(g), by employing ($u$, $v$, $w$) as the fractional coordinates of $k_a$, $k_b$ and $k_c$, all points as (0, $v$, $w$) and ($u$, 0, $w$) exhibit spin degenerate due to the specific little group of $^2m^\infty 1$[67]. Here, $^\infty 1$ is an invariant operation for each $k$ points, while $^2m$ combines a 180° spin rotation and a mirror operation on the fractional coordinates. On the other hand, $k$ points such as $B(u, v, 0)$, $E(u, v, 1/2)$ and $C(u, u, w)$ belong to a little group of $^11^\infty 1$, leading to nonrelativistic spin-splitting. The little group $^11^\infty 1$ is a fairly low symmetry, lacking any non-trivial operations that relate invariant subgroups of the two sublattices.

Importantly, the spin-polarization of band structure can be reversed due to the multiferroic nature of BiFeO$_3$, as illustrated in Fig. 2(c). When the ferroelectric polarization is switched by an electric field, the coupled atomic positions and magnetic moments effectively perform a *PT* operation on band structures. In consequence, as shown in Fig. 2(i) the band structure exhibits a reversal of spin polarization at each $k$ point.

## II. Fe$_2$Mo$_3$O$_8$

Another spin-splitting multiferroic Fe$_2$Mo$_3$O$_8$, as one of the typical type-I multiferroic, is recently reported an enhancement of magnetoelectric coupling by doping nonmagnetic Zn[68]. As demonstrated in Fig. 3(a-b), Fe$_2$Mo$_3$O$_8$ belongs to the magnetic space group of $P6_3'm'c$ with the hexagonal lattice constants of $a=b=5.77$ Å and $c=10.05$ Å[48]. The ferroelectric polarization of Fe$_2$Mo$_3$O$_8$ is 0.3 $\mu$C/cm$^2$ along $c$ axis with antiferromagnetic transition temperature $T_N$ around 60 K[69][70]. Three mirror planes, shown in the top view of Fe$_2$Mo$_3$O$_8$ in Fig. 3(a), are identified, with the corresponding symmetry operations on reciprocal $k$ vector $k_{a,b,c}$ and spin $S_{a,b,c}$ as follows:

$$TM_{[110]}(k_a, k_b, k_c; S_a, S_b, S_c) = (k_b, k_a, -k_c; -S_b, -S_a, S_c)$$

$$TM_a(k_a, k_b, k_c; S_a, S_b, S_c) = (k_a, -k_b, -k_c; -S_a, S_b, S_c)$$

$$TM_b(k_a, k_b, k_c; S_a, S_b, S_c) = (-k_a, k_b, -k_c; S_a, -S_b, S_c)$$

Similar to BiFeO$_3$, the spin-structure motif pairs illustrated in Fig. 3(c) imply the breaking of $PT\tau$, $U\tau$ and $P$ symmetries, while preserving a screw axis $6_{3c}$ that connects two opposite spin sublattices. At the $\Gamma$ point, its band structure is spin degenerate with the DOS identical for spin up and spin down electrons, both are essentials of altermagnets. In addition, the top view of the spin-structure motif pairs, shown in Fig. 3(e), are connected to each sublattice by $PT$ operation, resulting in the spin degenerate along the high symmetry paths, as illustrated in Fig. 3(f).

Although both BiFeO$_3$ and Fe$_2$Mo$_3$O$_8$ are spin degenerate along high symmetry paths, with spin splitting in other portions of Brillouin zone, their specific splitting behaviors are quite different. As demonstrated in Fig. 3(g), for Fe$_2$Mo$_3$O$_8$, the band structures at the surfaces of Brillouin zone are completely degenerate, that is due to the high symmetry little groups within these planes. By taking $B(u, v, 0)$, $C(u, u, w)$ and $E(u, v, 1/2)$ points as examples, the little groups of these $k$ points correspond to $^{-1}2^{\infty}1$, $^2m^{\infty}1$ and $^{-1}2^{\infty}1$ respectively, leading to two-fold symmetry in spin and thus no spin-splitting. Additionally, as shown in Fig. 3(h), spin splitting within the whole Brillouin zone exhibits a three-fold rotation symmetry on account of the screw axis $6_{3c}$ along $c$ axis. Finally, as shown in Fig. 3(i), when the ferroelectric polarization is switched, the spin-splitting contour map is also reversed due to the magnetoelectric coupling, corresponding to the transformation of the spin structure motifs from Fig. 3(c) to (d).

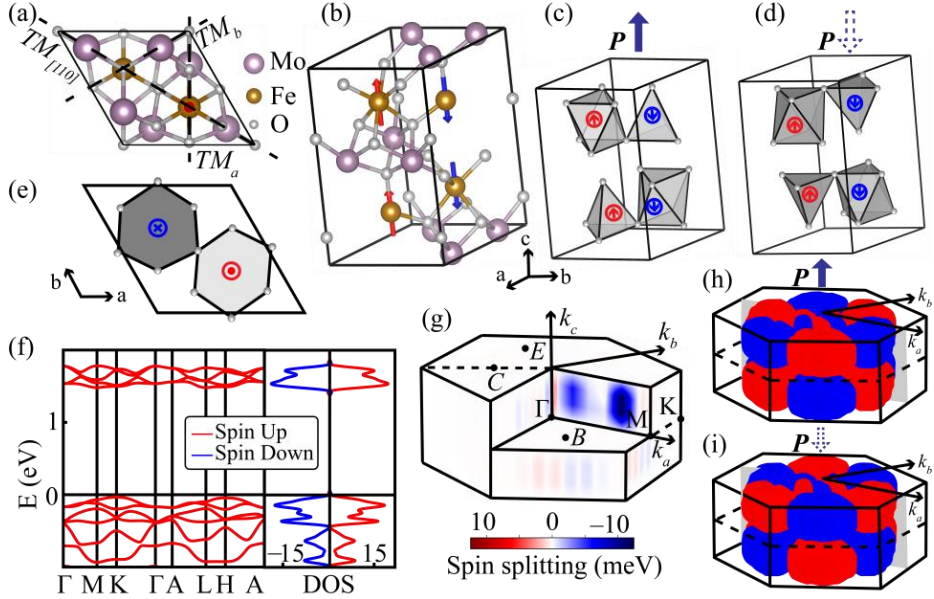

FIG. 3. (a) Top view of hexagonal $Fe_2Mo_3O_8$ in which dashed lines represent three mirror planes. (b) Atomic and spin structure of $Fe_2Mo_3O_8$. (c-d) Spin structure motifs of $Fe_2Mo_3O_8$ for up and down ferroelectric states. (e) Top view of (c). (f) Spin-resolved band structure along high-symmetry paths and DOS. (g) Spin-splitting energy dispersion of $Fe_2Mo_3O_8$. The splitting energy is calculated from the top valence band. (h-i) Contour map with the spin-splitting energy of 3 meV for two ferroelectric states of $Fe_2Mo_3O_8$.

## C. Multiferroic compensated ferrimagnet with zero net magnetization

The previous discussions are still valid when compensated ferrimagnets are considered. As shown in Fig. 4(a), as a typical multiferroic compensated ferrimagnet, $GaFeO_3$ with lattice constants of $a$=5.14 Å, $b$=8.82 Å and $c$=9.49 Å[71] has room temperature piezoelectricity[71], tunable translation temperature with the stoichiometry of oxygen and iron[73][71] . $GaFeO_3$ is an A-type compensated ferrimagnet with magnetic moments of -4.191 $\mu_B$ and 4.188 $\mu_B$ on two opposite Fe sublattices, -0.011 $\mu_B$ and 0.015 $\mu_B$ on Ga atoms, and -0.001 $\mu_B$ on Oxygen atoms, leading to net zero magnetic moment in unit cell[71][74]. The calculated band gap of GaFeO$_3$ is 2.27 eV, with a ferroelectric polarization of 24.9 μC/cm$^2$ along the *c* axis, which agrees well with previous results[75][76][72].

The principal symmetry operations of GaFeO$_3$, shown in Fig. 4(a-b), are as follows:

$$M_a\tau_{1/2abc}(k_a,k_b,k_c;S_a,S_b,S_c) = (-k_a+1/2, k_b+1/2, k_c+1/2; S_a, -S_b, S_c)$$

$$TC_{2c}\tau_{1/2c}(k_a,k_b,k_c;S_a,S_b,S_c) = (-k_a, -k_b, k_c+1/2; S_a, S_b, -S_c)$$

Where *M* represents the mirror plane, $TC_{2c}$ means a 180º rotation around *c* axis combined with time reversal operation. Thees symmetries together contribute GaFeO$_3$ to a zero net magnetic moment, with the magnetic space group of *Pna'2$_1$'*[77], and enable GaFeO$_3$ spin-polarized band structures due to the violation of *PTτ* and *Uτ* symmetry. However, we shall be aware that GaFeO$_3$ is actually a compensated ferrimagnet, not an antiferromagnet, leading to a distinct spin-splitting band structure compared to BiFeO$_3$ and Fe$_2$Mo$_3$O$_8$.

Without rotational symmetries connecting two magnetic sublattices, as shown in Fig. 4(d), multiferroic GaFeO$_3$ breaks spin degeneracy at the Γ point, where the spin motif pairs (from the perspective of [010] of GaFeO$_3$), break both *PTτ* and *Uτ* symmetries. As a result, as demonstrated in Fig. 4(e), spin-splitting occurs throughout the entire Brillouin Zone. The DOS is completely spin-polarized, confirming GaFeO$_3$ is a ferrimagnet. In addition, as shown in Fig. 4(e) and (f), the spin space group of $P^1n^1a^12_1^{\infty m}1$ with the little group at Γ point of $^1m^1m^12^{\infty m}1$ ensures spin-splitting at the Γ point. The notch in Fig. 4(f) is provided for better highlighting the spin-polarized high symmetry planes all over the Brillouin zone, and these symmetries are further illustrated in the splitting energy contour map in Fig. 4(g) with a mirror plane perpendicular to *a* axis and a 2-fold rotation operation around *c* axis. Moreover, Fig. 4(h), corresponding to the structure of Fig. 4(c), indicates the reversed spin-splitting under the reversal of ferroelectric polarization by an applied field.

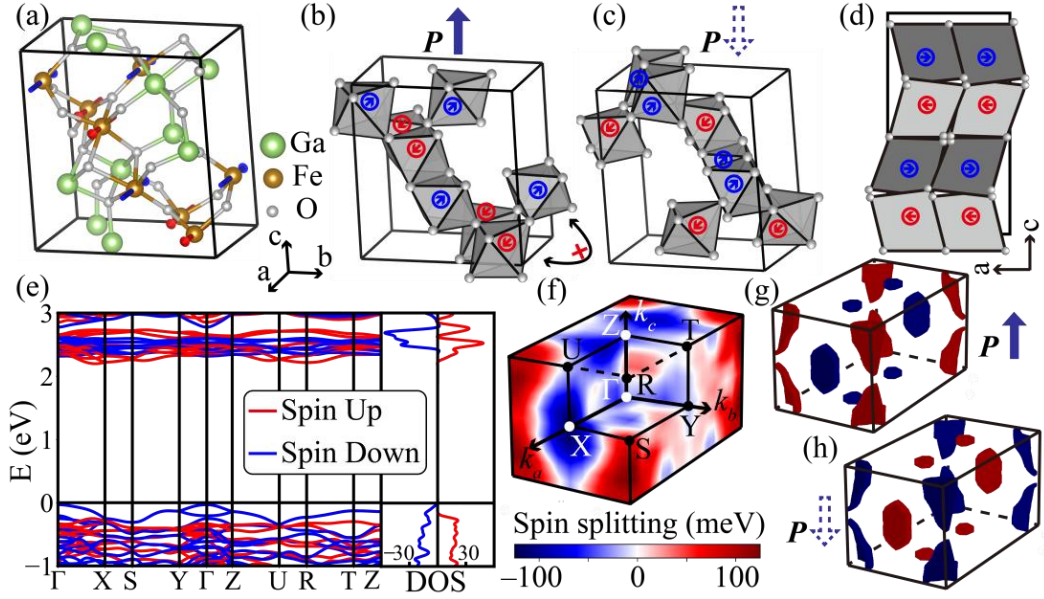

FIG. 4. (a) Atomic structure, (b-c) Spin structure motifs of the compensated ferrimagnet GaFeO$_3$ for two ferroelectric states. (d) The [010] perspective of (b). (e) Spin-resolved band structure along high-symmetry paths and DOS. (f) Spin-splitting energy dispersion for up ferroelectric state, and (g-h) 100 meV splitting energy contour maps for two ferroelectric states in the entire Brillouin zone.

## D. Potential applications

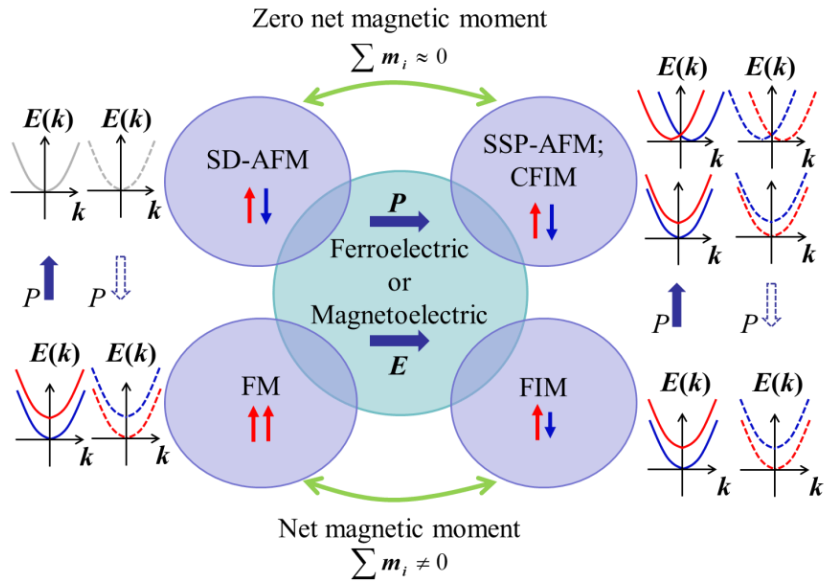

FIG. 5. Classification of multiferroic materials according to different type of magnetism. SD-AFM: spin

degenerate antiferromagnet; SSP-AFM: spin-splitting antiferromagnet; CFIM: compensated ferrimagnet; FM: ferromagnet; FIM: ferrimagnet. Red and blue arrows within each violet circle represent the magnetic moment directions on the sublattices of magnetic ions. Alongside each magnetic material, the schematic band structures are shown with red and blue curves representing two spin channels, where solid and dashed lines denote band structure before and after switching ferroelectric polarization $P$.

As shown in Fig. 5, by combing ferroelectric or magnetoelectric properties with magnetic orders, materials can now be explicitly classified into four categories. In ferroelectric or magnetoelectric SD-AFM, the reversal of polarization $P$ has no impact on electronic spin polarization. For ferroelectric FM and FIM with net magnetic moments, switchable spin splitting exists throughout the entire Brillouin Zone. A newly identified class of multiferroic antiferromagnet or compensated ferrimagnet, with zero net magnetic moments and spin-splitting band structures, provides a fascinating material platform that can be efficiently manipulated by an electric field. We name them as SSP-MAFM (spin-splitting multiferroic antiferromagnet) and MCFIM (multiferroic compensated ferrimagnet), with $BiFeO_3$, $Fe_2Mo_3O_8$, and $GaFeO_3$ serving as representative materials, as we have discussed previously.

Thanks to the coupling between ferroelectric polarization and magnetic moments, the spin-polarized band structures of SSP-MAFM and MCFIM can be reversed by electric field, offering potential perspective for designing functional structures and novel devices. For instance, with electric field switchable spin-polarization, SSP-MAFM and MCFIM may serve as tunnel barriers in multiferroic antiferromagnetic tunnel junctions (MAFTJs) with a structure of "NM(non-magnetic metal)/$BiFeO_3$ (SSP-MAFM or MCFIM)/$Mn_3Sn$ (metallic SSP-AFM)". The conductance of the whole MAFTJ would depend on the ferroelectric polarization as well as the magnetic states of SSP-AFM electrode, forming multilevel resistance states which may function as storage cells in high-density, energy-efficient, and external magnetic field robust antiferromagnetic memory. We leave the particular spin transport study in MAFTJs to our future work.

Moreover, in spin-splitting AFM and CFIM, the magnon dispersion qualitatively mirrors the splitting behavior of electron band structure[78], potentially lifting the magnon degeneracy (chirality of magnon) and enabling novel magnon spin transport phenomenon. Based on magnon splitting, for SSP-MAFM or MCFIM, the electric field control of magnon spin transport should also be feasible. Indeed, chiral magnon spin transport in BiFeO$_3$ has been successfully demonstrated recently[54] which implies SSP-MAFM or MCFIM may play a key role in future magnonic device applications.

**Summary**

In summary, we have investigated the design principles and elucidated the elusive combination of multiferroic properties and spin polarized antiferromagnet and compensated ferrimagnets with zero net magnetic moment. Simultaneous violation of *PTτ*, *Uτ* and *P* symmetries is essential for such antiferromagnets and ferrimagnets. To explicitly demonstrate different types of spin-splitting, we systematically analyzed three typical multiferroics including AFM BiFeO$_3$, Fe$_2$Mo$_3$O$_8$ and compensated ferrimagnet GaFeO$_3$.

All the studied materials are spin polarized in momentum space, while the specific spin-splitting varies according to their symmetry properties. BiFeO$_3$ and Fe$_2$Mo$_3$O$_8$ exhibit spin degenerate along high symmetry paths in the Brillouin Zone but show significant spin-splitting energies along lower-symmetry paths. Moreover, Fe$_2$Mo$_3$O$_8$, with higher symmetries, possesses more degenerate planes compared to BiFeO$_3$. In contrast, for multiferroic compensated ferrimagnet GaFeO$_3$ with zero net magnetization, the spin degeneracy is lifted over the whole Brillouin zone. Due to the coupling between ferroelectric and magnetic orders, the spin polarization of the band structures can be switched upon reversal of the ferroelectric polarization using an electric field. The discovery and identification of this new class of multiferroic materials enrich the existing family of multiferroic materials and offers valuable insights for the development of multifunctional spintronic devices based on electron and magnon spin transport phenomena with superior advantages.

**Acknowledgment**

This work was supported by the National Natural Science Foundation of China (grant No. T2394475, T2394470, 12174129).